\documentclass[sigconf]{acmart}
\usepackage[nolist]{acronym}
\usepackage{hyperref}
\usepackage{subfig}
\usepackage{tabularx}
\usepackage{xcolor}

\newcommand{\fix}[1]{\textcolor{black}{ #1}}

\AtBeginDocument{%
\providecommand\BibTeX{{%
\normalfont B\kern-0.5em{\scshape i\kern-0.25em b}\kern-0.8em\TeX}}}

\copyrightyear{2021} 
\acmYear{2021} 
\setcopyright{acmlicensed}\acmConference[AutomotiveUI '21]{13th International Conference on Automotive User Interfaces and Interactive Vehicular Applications}{September 9--14, 2021}{Leeds, United Kingdom}
\acmBooktitle{13th International Conference on Automotive User Interfaces and Interactive Vehicular Applications (AutomotiveUI '21), September 9--14, 2021, Leeds, United Kingdom}
\acmPrice{15.00}
\acmDOI{10.1145/3409118.3475140}
\acmISBN{978-1-4503-8063-8/21/09}

\begin{document}

\begin{acronym}[]
	\acro{ECU}{Electronic Control Unit}
	\acro{GDPR}{General Data Protection Regulation}
	\acro{HCI}{Human-Computer Interaction}
	\acro{HMI}{Human-Machine Interaction}
	\acro{HU}{Head Unit}
	\acro{IVIS}{In-Vehicle Information System}
	\acro{KPI}{Key Performance Indicator}
	\acro{OEM}{Original Equipment Manufacturer}
	\acro{OTA}{Over-The-Air}
	\acro{RH}{Research Hub}
	\acro{UCD}{User-centered Design}
	\acro{UX}{User Experience}
\end{acronym}

%
\title[Visualizing Event Sequence Data for User Behavior Evaluation of IVIS]{Visualizing Event Sequence Data for User Behavior Evaluation of In-Vehicle Information Systems}

%
\author{Patrick Ebel}
\email{ebel@cs.uni-koeln.de}
\orcid{0000-0002-4437-2821}
\affiliation{%
	\institution{University of Cologne}
	\city{Cologne}
	\country{Germany}}
	
\author{Christoph Lingenfelder}
\email{christoph.lingenfelder@daimler.com}
\affiliation{%
	\institution{MBition}
	\city{Berlin}
	\country{Germany}}

\author{Andreas Vogelsang}
\email{vogelsang@cs.uni-koeln.de}
\orcid{0000-0003-1041-0815}
\affiliation{%
	\institution{University of Cologne}
	\city{Cologne}
	\country{Germany}}

\renewcommand{\shortauthors}{Ebel et al.}
%

\begin{abstract}
With modern \acp{IVIS} becoming more capable and complex than ever, their evaluation becomes increasingly difficult. The analysis of large amounts of user behavior data can help to cope with this complexity and can support UX experts in designing \acp{IVIS} that serve customer needs and are safe to operate while driving. We, therefore, propose a Multi-level User Behavior Visualization Framework providing effective visualizations of user behavior data that is collected via telematics from production vehicles. Our approach visualizes user behavior data on three different levels: (1) The Task Level View aggregates event sequence data generated through touchscreen interactions to visualize user flows. (2) The Flow Level View allows comparing the individual flows based on a chosen metric. (3) The Sequence Level View provides detailed insights into touch interactions, glance, and driving behavior. Our case study proves that UX experts consider our approach a useful addition to their design process.
\end{abstract}

%
%


%
\maketitle

\section{Introduction}

Modern \acp{IVIS} are complex systems that offer a variety of features ranging from driving-related to infotainment functions with interaction options similar to those of smartphones and tablets. As technology progresses, so do the demands toward \ac{IVIS}, leading to customers expecting the same range of features and usability they are used to from their everyday digital devices. This makes it more challenging than ever to design automotive interfaces that are safe to use and still meet customer demands~\cite{Harvey.2016}. The introduction of large touchscreens as the main control interface further complicates this issue. Touchscreen interactions demand more visual attention than interfaces with tactile feedback~\cite{Pampel.2019}. They require users to visually verify that a correct selection has been made, making it necessary for drivers to take their eyes off the road. Since eyes-off-road durations longer than two seconds are proven to increase the crash risk~\cite{Klauer.2006}, the evaluation of touchscreen-based \acp{IVIS} also becomes a safety-related aspect apart from developing a system that satisfies the user needs in the best possible way. This added complexity makes it even harder to evaluate \acp{IVIS}, which is why UX experts require support from data-driven methods~\cite{Ebel.2020a}. 

The new generation of cars is more connected than ever and generates large amounts of data that cannot be sufficiently analyzed using traditional, mostly manual, approaches \cite{Agrawal.2015}. However, the analysis and visualization of big interaction data can significantly benefit user behavior evaluation~\cite{King.2017, Wei.2012} and offers great potential for the automotive domain~\cite{Orlovska.2018}. \citeauthor{Ebel.2020a}~\cite{Ebel.2020a} state that, currently, automotive interaction data is not used to its full potential. They describe that experts need aggregations of the large amounts of data and visualizations that allow deriving insights into user and driving behavior.

We propose a Multi-level User Behavior Visualization Framework for touch-based \acp{IVIS} consisting of three different levels of abstraction: (1) The task level that visualizes alternative interaction flows for one task (e.g., starting navigation), (2) the flow level that visualizes metrics of interest for the different interaction sequences of one flow (e.g., using the keyboard vs.\ using POIs to start navigation), and (3) the sequence level that augments single interaction sequences with contextual driving data such as speed or steering angle. UX experts can use the visualizations to effectively gain insights into user flows, their temporal differences, and the relation between user interactions, glance behavior, and driving behavior. \fix{This is not only valuable for current manual driving scenarios, but also for future driving scenarios, since, for example, the system could be used to evaluate the effect of secondary task engagement on take-over performance \cite{Eriksson.2017, Zeeb.2016, Du.2020}} In contrast to most of the related approaches, the data used in this work is collected and processed by a telematics-based big data processing framework that allows live data analysis on production vehicles. The presented visualizations were found very useful in an informal evaluation study with 4 automotive \ac{UX} experts.

\section{Background}
In this section, we discuss the current state-of-the-art in user behavior evaluation in the automotive industry and present different approaches on how to visualize user interactions and event sequence data in particular. Additionally, we introduce definitions that will be used throughout this work.

\subsection{User Behaviour Evaluation of Touchscreen-based IVISs}
Due to the high impact of digital solutions on the in-car \ac{UX} and the trend toward large touchscreens, being the current de facto standard control interface between driver and vehicle \cite{Harrington.2018}, the evaluation of touchscreen-based \ac{IVIS} gets increasingly important. A good \ac{UX} plays a major role for market success and is necessary to maintain competitiveness which makes usability evaluation of \acp{IVIS} a well-researched topic in the recent past \cite{Harvey.2011, Harvey.2016, Frison.2019b}. In contrast to the interaction with a smartphone or tablet, the interaction with \acp{IVIS} is only a secondary task. Since driving, still, is the primary task, the interaction with the touchscreen interface requires drivers to move their focus from the road toward the touchscreen. This focus shift has been shown to compromise safety and increase crash risk~\cite{Seppelt.2017}. Therefore, it is not only necessary to create a usable interface but also to assure that drivers are not overly distracted from the driving task when interacting with \acp{IVIS}. Assessing the driver's workload is still a challenging task and a variety of methods and data sources like physiological data, eye-tracking but also kinematic data are explored \cite{Palinko.2010, Schneegass.2013, Wollmer.2011, Risteska.2018, Kanaan.2019}. Multiple approaches tackle the task of predicting task completion times \cite{Schneegass.2011, Green.2015, Lee.2019, Kim.2014}, as well as visual demand \cite{Pettitt.2010, Large.2017, Pampel.2019, Purucker.2017} to assess, already in early development stages, how demanding the interaction with the in-vehicle touchscreen is.

However, most of the current approaches are based on questionnaires, explicit user observation, or performance-related measurements recorded during lab experiments or small-scale naturalistic driving studies. Additionally, most of the studies are designed to answer a specific research question that does not have a direct influence on the \acp{OEM} development and evaluation of \acp{IVIS}. \citeauthor{Lamm.2019}~\cite{Lamm.2019} describe that user behavior evaluations based on implicit data, generated from field usage, currently, do not play an important role in automotive UX development. On the other hand, \citeauthor{Ebel.2020a}~\cite{Ebel.2020a} found that automotive UX experts are in need of data-driven methods and visualizations that benefit a holistic system understanding based on data retrieved from production line vehicles. The authors argue that experts need tool support to understand what features are being used in which situations, how long it takes users to complete certain tasks, and how the interactions with \acp{IVIS} affect the driving behavior. Whereas first approaches address the potentials of big data analysis by incorporating telematics-powered evaluations \cite{Orlovska.2020b} they are still limited to naturalistic driving studies and especially the potentials of analyzing user interaction data are not yet well explored.

\subsection{Event Sequence Analysis}

Event sequence analysis is important for many domains ranging from web and software development \cite{Liu.2017, Wang.2016} to transportation \cite{Muntzinger.2010, Ebel.2020}. Event sequence data, being multiple series of timestamped events, ranges from website logs describing how users navigate the pages to energy flows showing how different types of energy are distributed within a city. Regardless of the particular use case, the main application is to compare different sequences of events (e.g. \textit{Homescreen} $\rightarrow$ \textit{Settings} $\rightarrow$ \textit{Privacy and Security}), their frequency (e.g. 35\% of users went from \textit{Settings} to \textit{Privacy and Security}), and the time intervals in between the events (e.g. it took them 5 seconds on average).

One group of event sequence visualizations is known as \textit{Sankey Diagrams} \cite{Friendly.2002, Riehmann.2005}. \textit{Sankey Diagrams} focus on the visualization of quantitative information of flows, their dependencies, and how they split in different paths. \textit{Sankey Diagrams} are directed graphs consisting of nodes and links. Each node represents a state in a flow and has weighted input and output links (except for source and sink nodes). Links represent transitions from a source node to a target node. The links' weight represents the flow quantity, visualized as the width of the respective link. Except for source nodes and sink nodes, the sum of incoming links equals the sum of the outgoing links. While being able to efficiently visualize flows between different nodes, originally \textit{Sankey Diagrams} do not take the temporal aspect of the transitions into consideration.
One approach that tackles the processing of temporal event data is presented by \citeauthor{Wongsuphasawat.2011}~\cite{Wongsuphasawat.2011} and is called \textit{LifeFlow}. The approach combines multiple event sequences into a tree while preserving the temporal spacing of events within a sequence. Whereas in \textit{LifeFlow} multiple event sequences are combined in a tree, \textit{OutFlow} \cite{Wongsuphasawat.2012} combines them into graphs, similar to Sankey diagrams. To represent the temporal spacing between events the authors introduce an additional type of edge whose width represents the duration of the transition. \textit{Sankey Diagrams}, \textit{LifeFlow}, and \textit{Outflow}, all focus on visualizing and analyzing the different flows, their distribution and their temporal aspects from one dataset. In contrast, the \textit{MatrixWave} approach presented by \citeauthor{Zhao.2015} aims to create a comparative analysis of multiple event sequence datasets by replacing the edge connector of the Sankey diagrams with transition matrices. Whereas the aforementioned approaches are solely focusing on visualizing event sequence data, other approaches aim to provide an overall framework for user behavior evaluation in a digital environment~\cite{Deka.2017}. \fix{In addition, commercial providers like UserTesting\footnote{https://www.usertesting.com}, UserZoom\footnote{https://www.userzoom.com} and alike offer tools to analyze user sequences. However, to meet the requirements of automotive UX experts, an approach has to be developed that allows to analyze event sequences on the one hand and provides direct insights into driving behavior and gaze behavior on the other hand.}
	\begin{figure*}
		\centering
		\includegraphics[width=0.8\linewidth]{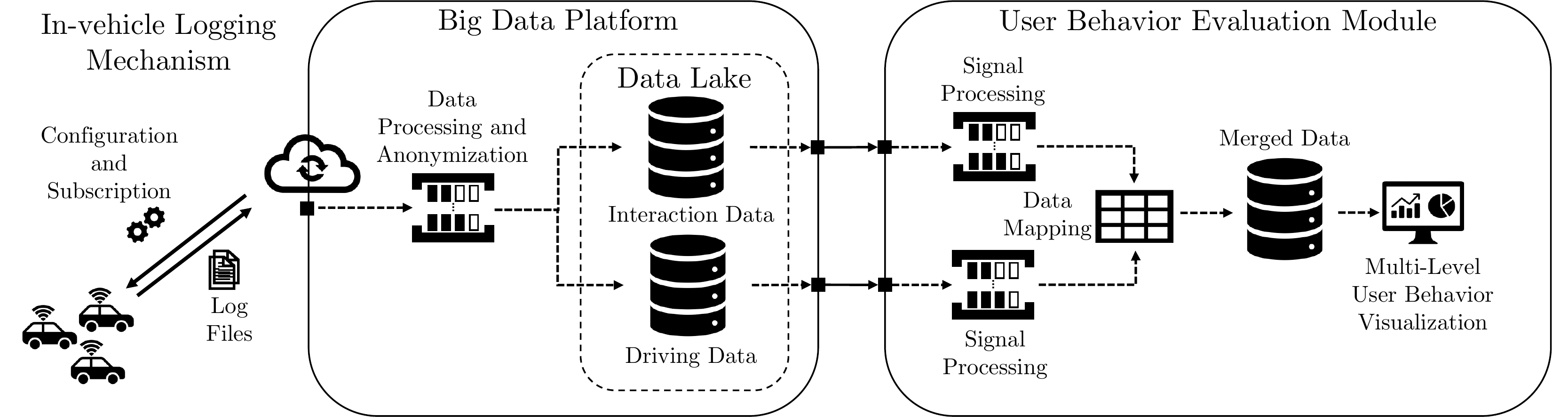}
		\caption{Architecture Overview} 
		\label{fig:Framework}
	\end{figure*}
\subsection{Definitions}
To create a common understanding in the further course of this work, the following definitions are introduced:

\textbf{Task.} A task is defined as an objective that a user must solve and consists of a defined start and end. The start and end of a task can further be defined by one or multiple conditions, being for example certain UI elements. A task can consist of multiple flows, meaning that the progression on how a user went from the start to the end is arbitrary. \fix{\textit{Example: ``Starting in the map view: Start the navigation to any destination.''}}

\textbf{User Flow/Path.} A user flow/path describes a linear series of events performed by a user to complete a certain task. \fix{\textit{Example: NavigateToButton\_tap $\rightarrow$ OnScreenKeyboard\_tap $\rightarrow$ List\_tap $\rightarrow$ StartNavigationButton\_tap}}

\textbf{Sequence.} A sequence is defined as a specific series of timestamped interactions performed by one user. \fix{\textit{Example: [(NavigateToButton, tap, timestamp, session\_id), (OnScreenKeyboard, timestamp, session\_id), (List, tap, timestamp, session\_id), (StartNavigationButton, tap, timestamp, session\_id)]}}

\textbf{Event.} An event is a specific user interaction defined by the triggered UI element, the gesture type, and its timestamp. \fix{\textit{Example: (NavigateToButton, tap, timestamp, session\_id)}}

\section{Multi-Level User Behavior Visualization}

Our approach benefits a holistic user behavior evaluation of \acp{IVIS} by visualizing different levels of abstraction of user/system interaction with the touchscreen interface. We have designed these three different visualizations as each of them satisfies certain requirements of UX experts \cite{Ebel.2020a}. The Task Level View allows UX experts to inspect how users navigate within the system, what the main interaction flows are, and how they relate to each other. The Flow Level View provides a quantitative comparison of different flows based on a chosen metric. Finally, the Sequence Level View enables UX experts to analyze certain sequences regarding the interrelation between touch interactions, glance behavior, and driving parameters. 

The data used in this work is collected from production vehicles without a specifically designed test environment or a defined group of participants. This, in theory, enables the data to be collected from every modern car in the fleet of our research partner, a leading German \ac{OEM}. The usage of natural user interaction data has three main advantages compared to data retrieved from lab experiments: (1) A large amount of data can be collected from the whole user base; (2) No specific costs for controlled experiments are incurred; (3) The context of use i.e. the driving situation is inherently contained in the data.

In the following, the data collection and processing framework is introduced followed by a detailed description of the aforementioned visualizations. 

\subsection{Telematics Architecture}

The data collection and processing is based on a feature-usage logging mechanism for the telematics and infotainment system. It enables \ac{OTA} data transfer to the Big Data Platform where the data is processed and off-board data analytics are performed to generate insights into user behavior with the \acp{IVIS}. The system architecture consists of three major parts: (1) the \textit{In-vehicle Logging Mechanism}, (2) the \textit{Big Data Platform}, and (3) the \textit{User Behavior Evaluation Module} itself. An overview of the system is given in Figure \ref{fig:Framework}.

The In-vehicle Logging Mechanism collects user interaction data from the \ac{HMI} interface and driving-related data from the vehicle bus. At the beginning of a trip, each car sends a request to the Big Data Platform, asking if a new configuration file is available, and gets assigned a session ID. Since no personal data is transmitted, the session ID is the only identifier linking the datapoints of a trip. Afterward, data packages containing log files are sent to the Big Data Platform in regular intervals until the ignition is switched off. The Big Data Platform receives, processes, further anonymizes (e.g. altering the timestamps), and stores the data in a data lake. 


The User Behavior Evaluation Module, developed in the course of this work, then accesses the user interaction data (event sequence data) and the driving data stored in the datalake. The signals are processed and the driving data is merged with the interaction data using the session ID. Since this system is already available in the production line vehicles of our research partner, it was not necessary for us to add further instrumentation.

\subsection{Data Collection and Processing}
The visualizations shown in this paper are based on data from 27,787 sessions generated by 493 individual test vehicles collected through the introduced telematics logging system. \fix{The vehicles are used for a diverse range of internal testing procedures of our research partner. No special selection criteria were applied and therefore all vehicles with the most recent telematic architecture contributed to the data collection.} The event sequence data consists of timestamped events containing the name of the interactive UI element that was triggered by the user and the type of gesture that was detected. First, all event sequences that satisfy the start and end condition (e.g. the respective UI elements) of a task and do not meet a task-specific termination criterion are extracted and are assigned a \textit{Task ID}. The termination criterion is intended to give users the ability to customize the evaluations to meet their needs. It can be defined as a set of specific UI elements or a maximum time limit $t_{max}$ that applies to the interval between two interactions. All sequences in which the termination criterion is met will be cleansed. If, for example, it is defined that a maximum of 60 seconds $t_{max}=60$ may elapse between two events and otherwise the task is considered incomplete, all sequences in which this applies will be cleansed. After sequence extraction, every sequence is assigned a unique \textit{Sequence ID} and all sequences that consist of the same ordered list of events are assigned the same \textit{Flow ID}. The driving data used in this work (steering wheel angle and vehicle speed), is extracted at a frequency of 5\,Hz and parsed to a human-readable format. The glance data is continuously collected using a face-facing camera located behind the steering wheel. The driver's field of view is divided into different regions of interest such that datapoint consists of start and end time, and region ID of a glance. If the region in the driver's focus changes, a new datapoint is collected. Since the data is processed in the vehicle no video data is transferred at any time.

\subsection{Task Level View}

	\begin{figure}
		\centering
		\includegraphics[width=\linewidth]{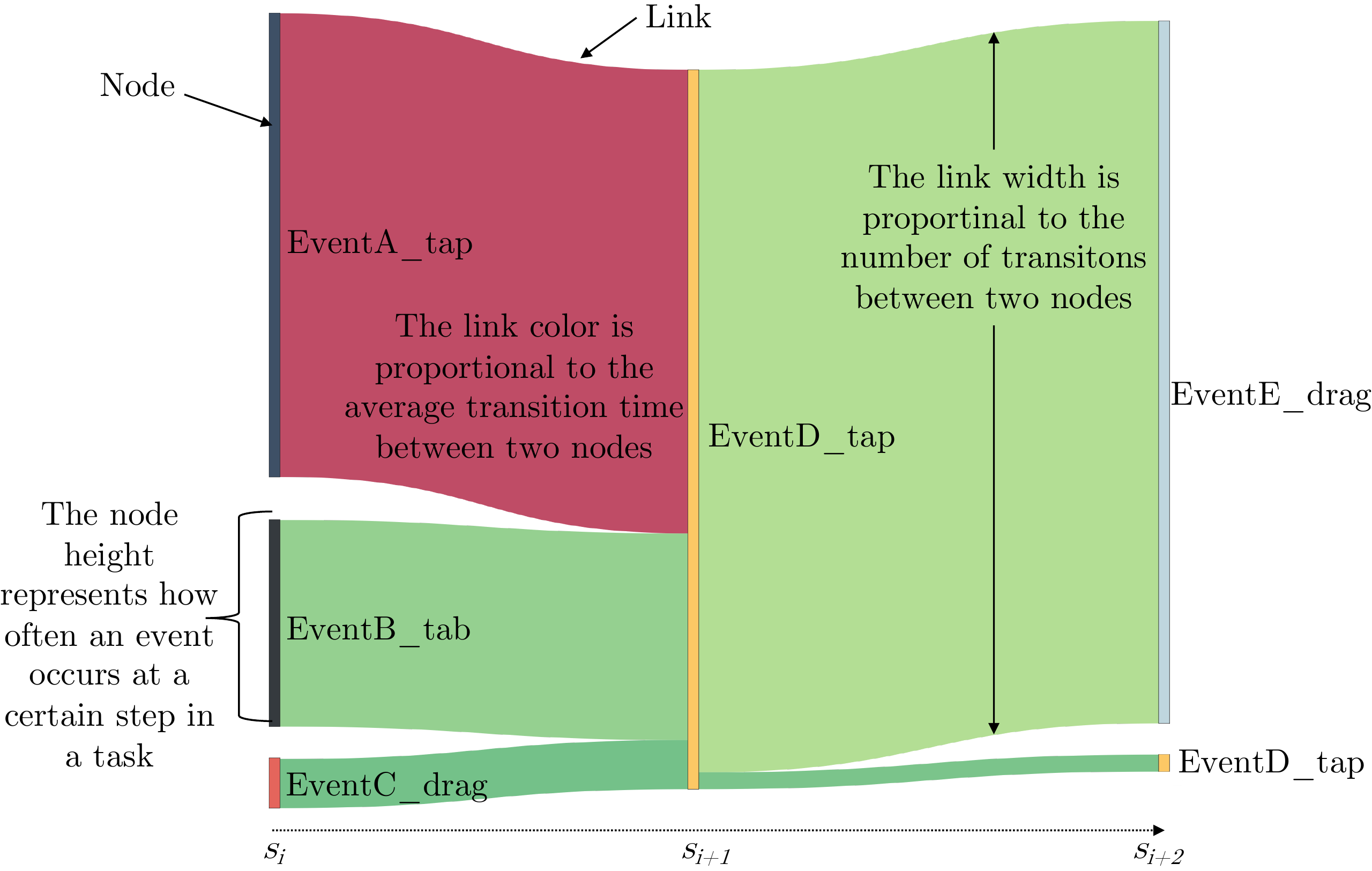} 
		\caption{Visual Encoding of Nodes and Links as used in the Task Level View} 
		\label{fig:SankeyAnnotations} 
	\end{figure}
	
The Task Level View visualizes how users navigate within the system to fulfill a certain task. Event sequence data, generated through touchscreen interactions, is aggregated and visualized in form of an adapted Sankey diagram. We decided to choose Sankey diagrams as the basis for the Task Level View because of their popularity and their efficient way to visualize multiple different flows and their distribution. We address the main weakness of Sankey diagrams, being that they do not encode temporal information by introducing color-coded links. Being able to see the most frequent user flows and their temporal attributes at one glance assists UX experts in identifying unintended or unexpected user behavior. The individual components and their visual encoding are shown in Figure \ref{fig:SankeyAnnotations}.

\textbf{Nodes.} Each Node represents an event at a certain step in a task. The nodes are visualized as rectangles whose height is proportional to the event's cardinality at a certain step in the task. The name of the UI element and the gesture (annotated as \textit{\_tap, \_drag or \_other}) used for an interaction are displayed next to the Node (see Figure \ref{fig:SankeyAnnotations}. The horizontal position indicates the step in the flow at which the event happened. Thus, nodes that are vertically aligned represent events at the same step in a task. In Figure \ref{fig:SankeyAnnotations}, step $s_i$ comprises three different events, whereas $s_{i+1}$ comprises only one event, meaning that whatever users did in $s_1$, they all made the same interaction (\textit{EventD\_tap}) in $s_{i+1}$. Nodes that represent the same event at different steps are colored equally (compare \textit{EventD\_tap} in Figure \ref{fig:SankeyAnnotations}). When hovering over a node, the number of entities, incoming, and outgoing links are displayed.

\textbf{Links.} Each link connects two nodes and therefore represents a transition between two events. The link's width is proportional to the number of transitions between the source node and the target node. The link color represents the average transition time between two events. The time is normalized to [0,1] using min-max-normalization, with higher values representing slower transitions. The normalized values are mapped to a linear color scale from green (0; short time) to red (1; long time). As displayed in Figure ref{fig:SankeyAnnotations}, the transition \textit{EventA\_tab} $\rightarrow$ \textit{EventD\_tab} is the most prominent one moving from $s_i$ to $s_{i+1}$ but also the slowest. When hovering over a link, a description is given describing in how many sequences (absolute and relative values) users went from the source node to the target node and how much time it took on average.

To create a visualization, the events that indicate the start and the end of a task need to be defined. The optional parameter $p_{min}$ allows users to set a lower bound, such that only flows with a relative frequency greater than $p_{min}$ are displayed. This filter increases readability since Sankey diagrams are hard to read for a large number of nodes \cite{Zhao.2015}. Additionally, UX experts can define a set of interactions that are represented as a single node even if they occur multiple times in succession (e.g. keyboard taps).

	\begin{figure}
		\centering
		\includegraphics[width=\linewidth]{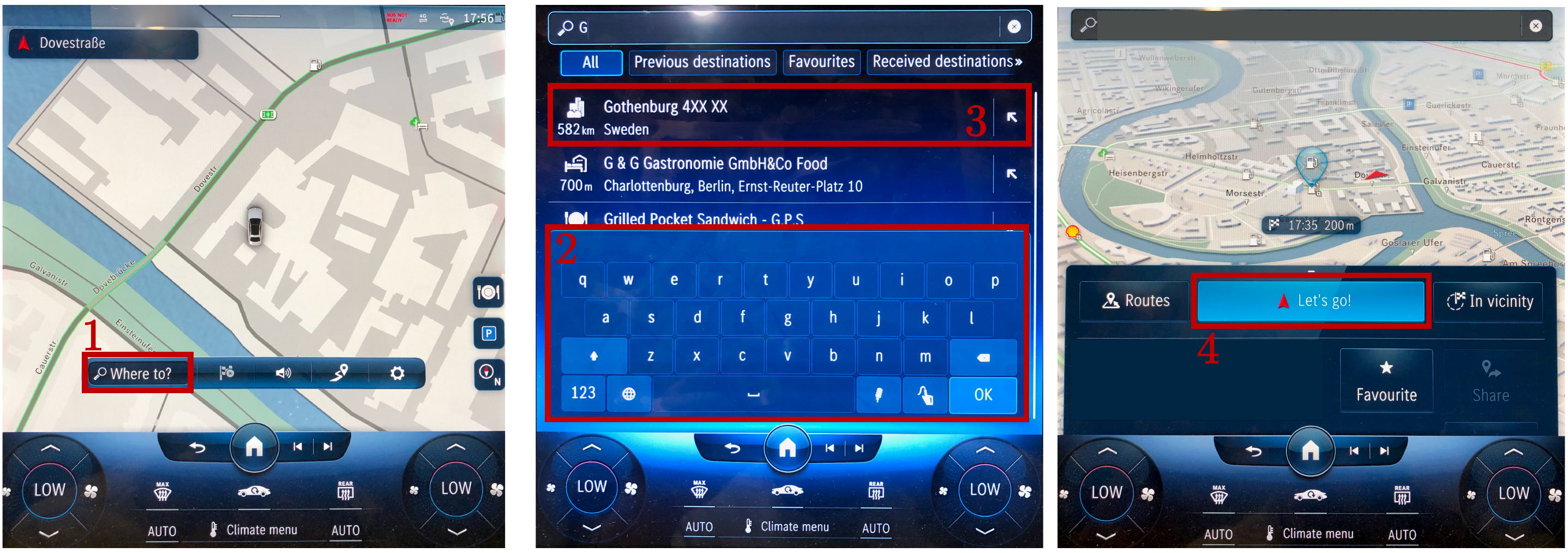} 
		\caption{Example of Flow 1: \textit{NavigateToButton\_tap $\rightarrow$ OnScreenKeyboard\_tap $\rightarrow$ List\_tap $\rightarrow$ StartNavigationButton\_tap}} 
		\label{fig:Screens} 
	\end{figure}
	
	\begin{figure*}
		\centering
		\includegraphics[width=\linewidth]{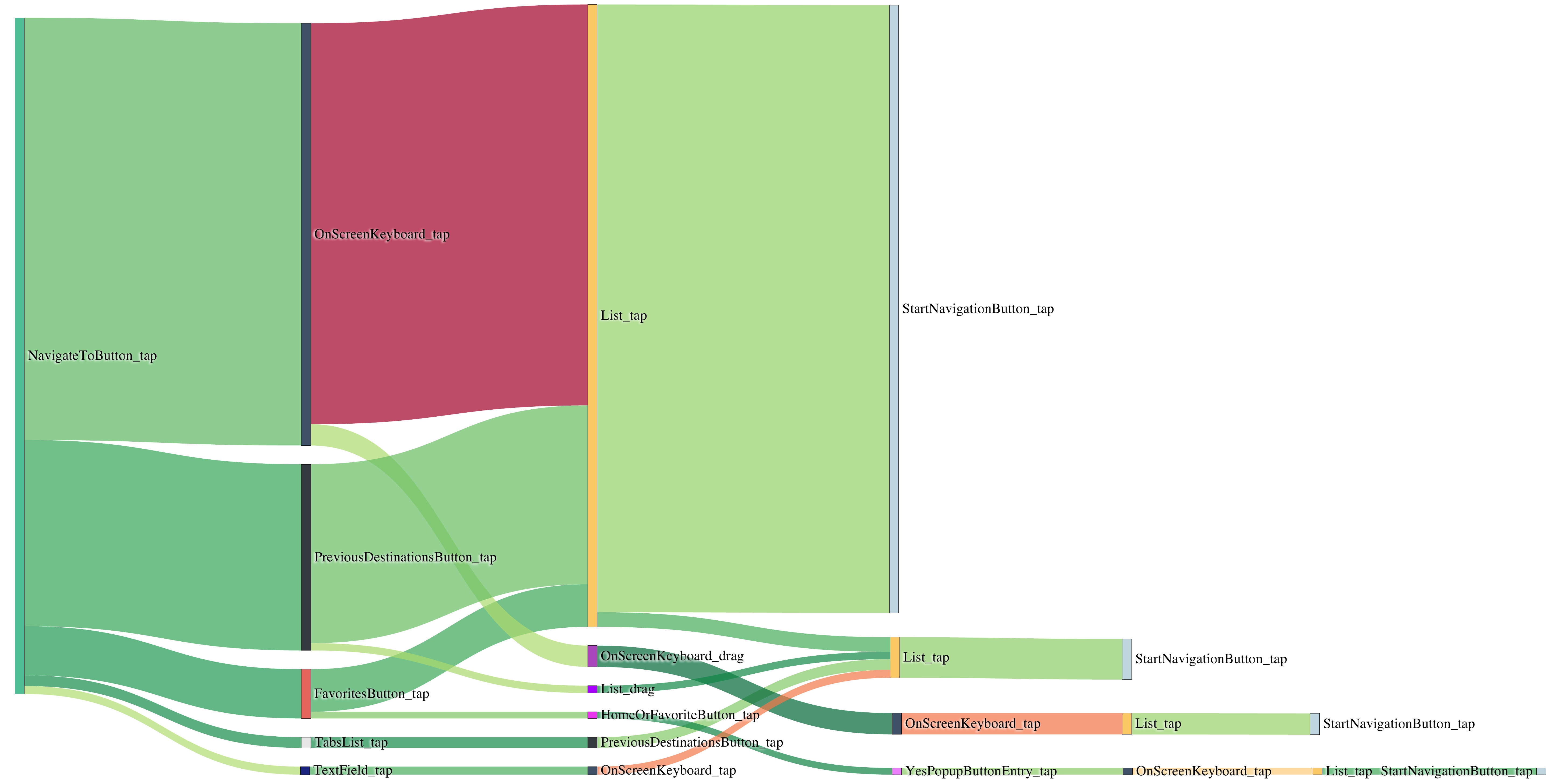} 
		\caption{Task Level View ($t_{max}=60s, p_{min}=0.005$)} 
		\label{fig:SankeyNavi} 
	\end{figure*}

\textbf{Example.} Figure \ref{fig:SankeyNavi} shows the Task Level View for a navigation task that starts with opening the navigation app from the map view on the Homescreen (\textit{NavigateToButton\_tap}) and ends with confirming that the route guidance shall be started (\textit{StartNavigateButton\_tap}). Investigating the different flows, one can clearly see that, whereas in most of the cases users directly started to use the keyboard to enter their destination (62\%), some users chose to use the option to select a destination out of their previous destinations (28\%) or their pre-entered favorites (7\%). After typing on the keyboard (\textit{OnScreenKeyboard\_tap}) to enter the destination, the majority of users directly chose an element out of the list of suggested destinations presented by the system (\textit{List\_tap}). Afterward, the majority then started the route guidance by accepting the proposed route (\textit{StartNavigateButton\_tap}). An example of this flow and how it looks like in the production vehicles \ac{IVIS} is given in Figure \ref{fig:Screens}. Apart from identifying the most popular flows, the Task Level View also assists UX experts in finding unintended user behavior. For example, after the first interaction (\textit{NavigateToButton\_tap}) the keyboard automatically opens and users can directly start typing. However, roughly one percent of the users first clicked on the text field and the started typing. This could lead to the hypothesis that users did not anticipate that the text field is already pre-selected and that they therefore tried to activate it by clicking on it. 

Apart from visualizing certain user flows and their popularity, the color-coding of the links allows conclusions to be drawn about interaction times. Typing on the keyboard (\textit{OnScreenKeyboard\_tap}) is by far the most time-consuming interaction in the presented task. Since it is the only aggregated event consisting of multiple user interactions this information may not be surprising. It nevertheless shows that a large portion of the time on task can be attributed to typing on the keyboard. Taking a closer look at the second step of the task, one can observe that users need about 2.3 seconds to choose a destination out of a list of pre-entered favorites (\textit{FavoritesButton\_tap}) is, whereas they need roughly 3 seconds to choose a destination out of a list containing all previous destinations (\textit{PreviousDestinationsButton\_tap}). This difference could be attributed to the fact that the favorites list is a structured list that tends to have fewer entries than the chronologically sorted list of previous destinations.

\subsection{Flow Level View}

	\begin{figure*}
		\centering
		\includegraphics[width=\linewidth]{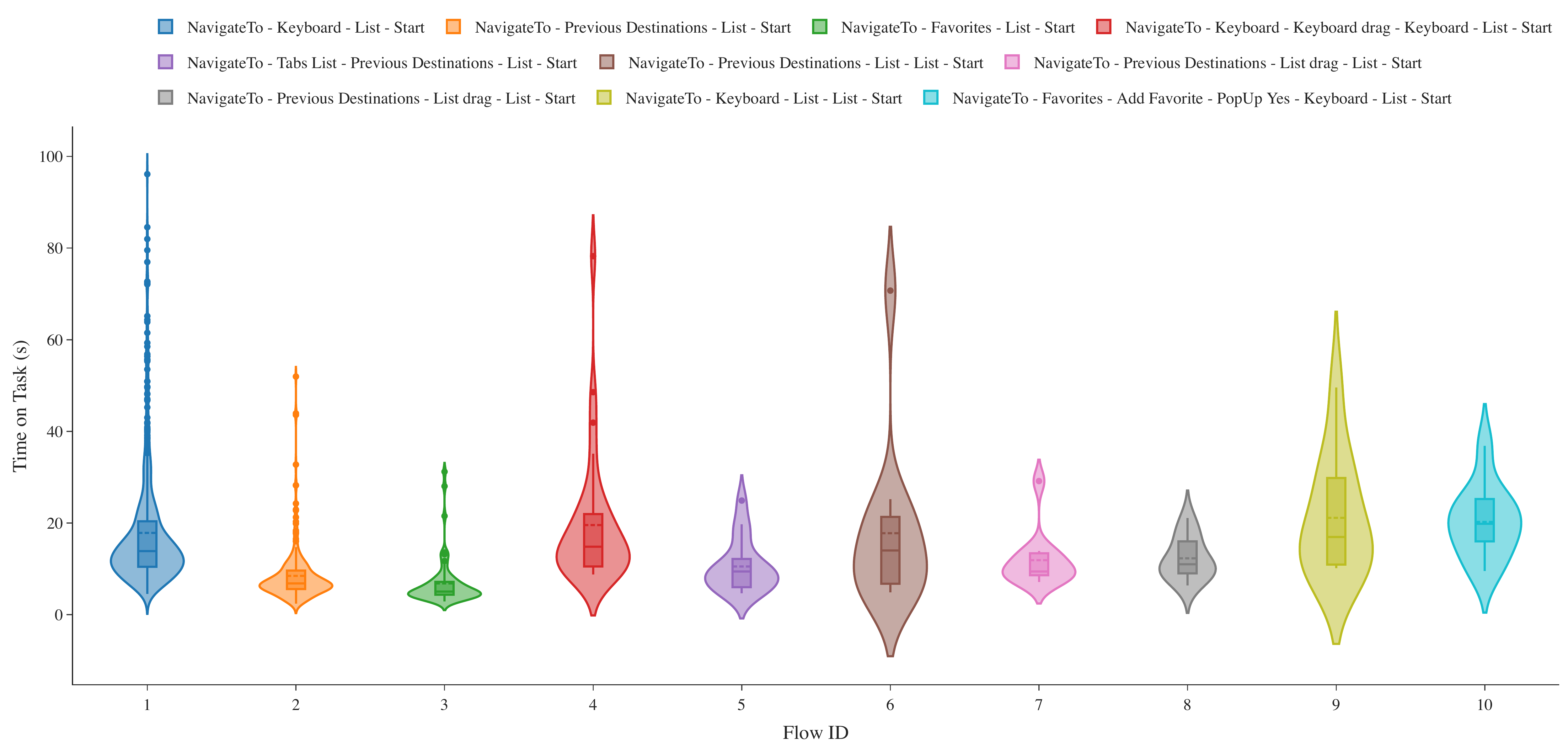} 
		\caption{Flow Level View (The names of the events have been shortened)} 
		\label{fig:ViolinNavi} 
	\end{figure*}

Whereas the Task Level View provides an overview of the different flows and their proportion, other metrics like for example the \textit{time on task} of specific flows and how they compare are not sufficiently visualized. The Flow Level View (Figure \ref{fig:ViolinNavi}) addresses this shortcoming by visualizing the distribution of a certain metric (for the example we use the time on task) of all sequences that belong to a flow (see Figure \ref{fig:ViolinNavi}). By visualizing the time on task as violin plots, two main insights can be generated. On the one hand, multiple statistics (e.g. min/max, mean, interquartile range) are visualized when hovering over the plot. UX experts can assess the displayed metrics and compare them to target values or industry guidelines \cite{Green.1999, SAE.2004}. On the other hand, displaying the violin plots next to each other allows a visual comparison of the individual flows. For example when comparing the distribution of flow 1 (\textit{NavigateToButton\_tap $\rightarrow$ OnScreenKeyboard\_tap $\rightarrow$ List\_tap $\rightarrow$ StartNavigationButton\_tap}), flow 2 (\textit{NavigateToButton\_tap $\rightarrow$ PreviousDestinationsButton\_tap $\rightarrow$ List\_tap $\rightarrow$ StartNavigationButton\_tap}), and flow 3 (\textit{NavigateToButton\_tap $\rightarrow$ FavoritesButton\_tap $\rightarrow$ List\_tap $\rightarrow$ StartNavigationButton\_tap}) one can observe that the time on task when using the keyboard is nearly double the time needed compared to either using the favorite or previous destination options. Comparing the latter (flow 2 and flow 3), using the favorites option is about two seconds faster than using the previous destination option. Whereas this difference has already been identified in the example describing the Task Level View, the impact on the whole task completion time can now be quantified. 
	
\subsection{Sequence Level View}

	\begin{figure*}
		\centering
		\subfloat[Flow 8: \textit{NavigateToButton\_tap $\rightarrow$ OnScreenKeyboard\_tap $\rightarrow$ List\_drag $\rightarrow$ List\_tap $\rightarrow$ StartNavigationButton\_tap}]{\includegraphics[width=\linewidth]{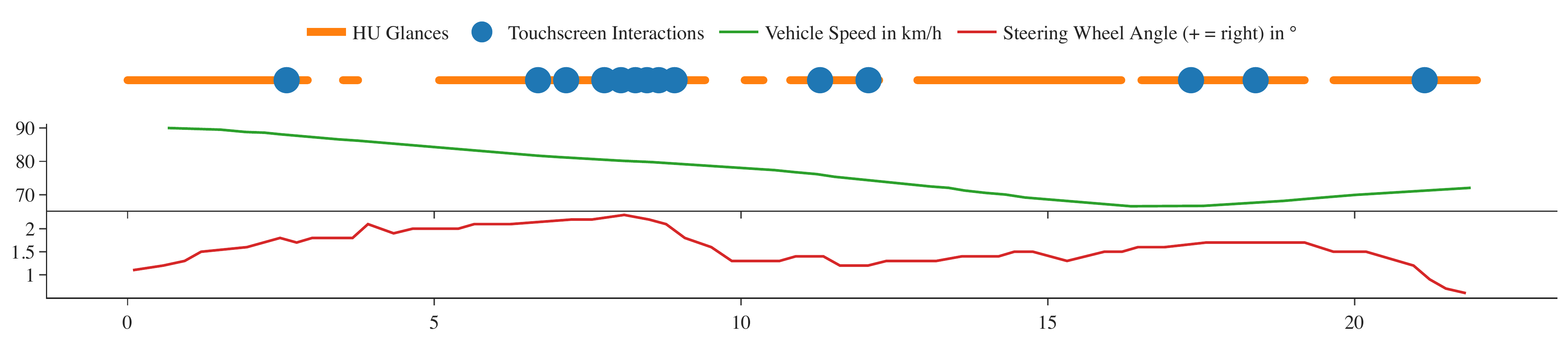}\label{fig:SequencePlot1}}
		\newline
        \subfloat[Flow 2: \textit{NavigateToButton\_tap $\rightarrow$ PreviousDestinationsButton\_tap $\rightarrow$ List\_tap $\rightarrow$ StartNavigationButton\_tap}]{\includegraphics[width=\linewidth]{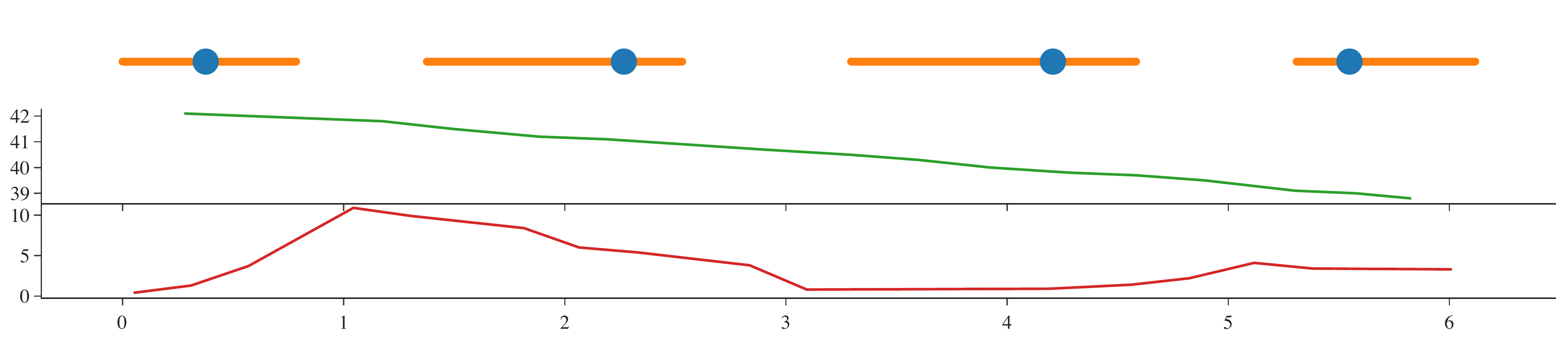}\label{fig:SequencePlot2}}
        \newline
        \subfloat[Flow 6: \textit{NavigateToButton\_tap $\rightarrow$ TextField\_tap $\rightarrow$ OnScreenKeyboard\_tap $\rightarrow$ List\_tap $\rightarrow$ StartNavigationButton\_tap}]{\includegraphics[width=\linewidth]{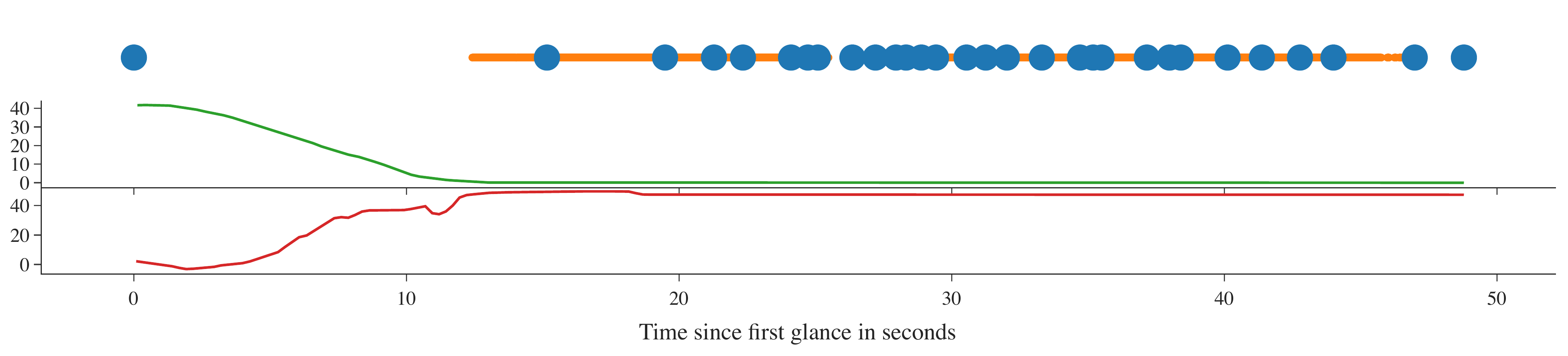}\label{fig:SequencePlot3}}
		\caption{Sequence Level View}
		\label{fig:SequencePlot} 
	\end{figure*}
	
An increased visual distraction from the driving task toward non-driving-related tasks is associated with increased crash risk~\cite{Green.2000, Lee.2008}. Thus, insights into the interrelation of user interactions, glance behavior, and driving behavior can yield valuable information for UX experts regarding the safety assessment of touch-based \acp{IVIS}. Whereas the previous views visualize general trends, the proposed Sequence Level View (see Figure \ref{fig:SequencePlot}) generates such insights by making it easy to identify long off-road glances, demanding click patterns, or other safety-critical driving behavior. 

The visualization consists of two main parts: The upper part is an overlay of touchscreen interactions (blue dots) and the driver's glances toward the center display (orange lines). Each dot represents one interaction and each line indicates the duration of a glance toward the display. The lower visualization, consisting of two graphs, represents the driving-related data (vehicle speed (green line) steering wheel angle (red line)). In Figure \ref{fig:SequencePlot}, three different sequences are visualized, emphasizing the importance to set the evaluation of user flows in perspective to the context.

In Figure \ref{fig:SequencePlot1} a specific sequence of Flow 8 is visualized. One can observe that it took the driver five long glances ($t>2\,\text{s}$) and three short glances ($t<2\,\text{s}$) to fulfill a task of 14 interactions whereas 10 of the interactions are keyboard interactions. Additionally, we can observe that the vehicle speed decreased after starting to type on the display and increased again at the end of the sequence. The change in the steering wheel angle is generally low, however, one can detect a small drift during the first intense typing interaction and a small correction after the second long glance. Whereas the first sequence took around 20 seconds for completion, the sequence using the previous destination option only took roughly six seconds, requiring four glances and four interactions. The vehicle speed did only slightly decrease during the interaction. In contrast to the two above sequences, the sequence displayed in Figure \ref{fig:SequencePlot3} consists of 30 touch interactions (25 keyboard interactions) but only two glances. During normal driving, taking the eyes off the road for such a long period of time would be considered highly safety-critical. However, considering the vehicle speed and the steering wheel angle, one can conclude that the driver pulled over to the right and stopped the car before starting to interact with the HMI. Therefore, this is not considered critical behavior and shows that certain statistical outliers need to be assessed individually.

\section{Informal Evaluation}
To assess the usefulness of the proposed approach and to answer the question of whether the visualizations are suited to generate knowledge from large amounts of event sequence data, we conducted a user study. The goal of the study was to understand how participants interact with the presented visualizations when trying to answer questions regarding user behavior. Therefore, we recruited four automotive UX experts (P1-P4, one UX Researcher, and three UX Designers with \fix{3, 9, 4, and 18 years of working experience respectively}) from our research partner. Two participants \fix{were directly involved in the design and development of the HMI} analyzed in this study. The examples presented in the previous sections were sent to the participants as an interactive web page and a document containing further information regarding the presented interface was provided. Due to the ongoing Covid-19 pandemic, we conducted the interviews remotely using Zoom. During the study, the participants were asked to share their screen and the interviews were recorded using the built-in audio and video recorder. Each interview comprised an introduction (20 minutes), an interactive part (30 minutes), and a discussion (10 minutes). During the introduction, we presented the objective of the presented system, the telematics framework, the exemplary task (screenshots and the respective UI elements), and demonstrated the features of the system. We asked the participants to explore the different visualizations and to ask questions in case some explanations were unclear. During the interactive part, the participants were asked to answer a list of seven distinct study questions (see Table \ref{tab:tasks}). The questions are inspired by the needs and potentials identified in~\cite{Ebel.2020a} and aim to test if the visualizations are suited to generate the anticipated insights.

After interacting with the visualizations to answer the study questions, the participants were given another 10 minutes to explore the visualizations to find any behavioral patterns that might indicate usability issues. After the interactive part, we initiated a discussion regarding the different visualizations and how the participants might integrate them into their design process. After the interview, the participants were asked to answer a survey with 8 questions addressing the usefulness of the system and its potentials with regard to their workflow. The questions demanded answers on a 7-point Likert scale ranging from strongly disagree (1) to strongly agree (7).\footnote{Questions and results are given here: \url{https://doi.org/10.6084/m9.figshare.14915550}}

\subsection{Generated User Behavior Insights}
	\begin{table*}
		\caption{Study Questions and Objectives.}
		\centering
		\label{tab:tasks}
		\small
		\begin{tabular}{@{}clll@{}}
			\toprule
		    \# & View Level & Question & Objective\\
			\midrule
			SQ1 & Task & Which path do most users take to start the navigation? & \textit{Traverse graph and interpret link width}\\
			SQ2 & Task & Do users prefer the favorites or previous destination option? & \textit{Interpret node height} \\
			SQ3 & Task & Which interaction is the most time-consuming? & \textit{Interpret link color}\\
			SQ4 & Flow & What is the fastest way to start the navigation? & \textit{Interpret metrics shown as hovering elements}\\
			SQ5 & Flow & Which flows are interesting to compare and why? & \textit{Compare distributions to find distinctive features}\\
			SQ6 & Sequence & Can you observe any safety-critical behavior? & \textit{Interpret glance duration and click behavior}\\
			SQ7 & Sequence & How do you interpret the driving situation? & \textit{Interpret driving parameters }\\
			\bottomrule
		\end{tabular}
	\end{table*}
In the following, we assess whether the visualizations are suited to answer the study questions (see Table \ref{tab:tasks}).

\textbf{Task Level View.}
All four participants answered the questions regarding the Task Level View (SQ1-SQ3) without additional support. They compared the respective links and nodes to answer SQ1 and SQ2 and interpreted the color coding as intended to find the most time-consuming interaction (SQ3). Also, P1 and P2 were particularly interested in flow 4: ``I can easily see that most people use our system as intended and I'm not overly concerned with flows that only occur very few times. But seeing 5 percent using the drag gesture on a keyboard [...] I would like to get into more detail'' (P1). Interestingly, during the interviews, we observed that the participants used the Task Level View as a kind of reference. Often when an anomaly or a pattern of interest was detected in one of the other views, participants invoked the Task Level View to verify what role the flow or the specific interaction plays in the overall context of the task.

\textbf{Flow Level View.}
Compared to the Task Level View, only two participants (P1 and P3) answered SQ4 without any further information. They quickly decided to base their answer on the median time on task and therefore identified flow 3 to be the fastest way to start the navigation. Whereas P1 and P3 were familiar with boxplots and violin plots, this kind of visualization was unknown for P2 and P4. P4 stated that \textit{``[he] would need to get more familiar with this kind of statistics''}. They, therefore, needed some additional assistance, but then solved SQ4 in similar a manner as P1 and P3 did. P1 adds that: \textit{``Interpreting this visualization gets easier the more often one uses it in the daily work''}. When asked to compare flows that might yield interesting insights, P1 argued that the distribution of the time on task could be used as a complexity measure that a more widespread distribution could indicate a more complex flow. Therefore, \fix{the interviewee} compared flow 1 and flow 6, with the only difference being that in flow 6 people clicked in the text field before they started typing. Based on the more widespread distribution of flow 6, \fix{P1} argued that \textit{``some people seem to have difficulties in understanding that the text field is already activated and that there is no need to tap on it. This seems to lead to longer interaction times''}.

\textbf{Sequence Level View.}
Working with the three different examples of the Sequence Level View (SQ6 and SQ7) all participants were able to derive certain hypotheses regarding driver distraction based on the glance and driving behavior. All participants found that the glances in Figure \ref{fig:SequencePlot1} are critically long. Regarding the long glance without any interaction after typing on the keyboard, P4 states that it \textit{``[...] might be due to a slow internet connection or because the intended destination was not in the list of suggestions''}. Based on the vehicle speed and the steering wheel angle participants concluded that the person was distracted by the interaction and the long glances. P1 explains that \textit{``[d]uring the keyboard interaction, there is an increasing deviation in the steering angle and a correction at the end of the interaction, even though it may be small in absolute terms''}. In contrast to Figure \ref{fig:SequencePlot1}, the glances in Figure \ref{fig:SequencePlot2} were considered not critical by all participants. P1 remarks \textit{``[t]hat's one glance per interaction, just like we want it to be''} and further explains that \fix{one} cannot attribute the deviation of the steering angle to the interaction with the HU. P3 was particularly interested in why people are in need to focus on the head unit after interacting with it and suspects that users want to have visual feedback on their interaction. Regarding the sequence visualized in Figure \ref{fig:SequencePlot3} all participants quickly identified that the driver pulled over to the right and then started engaging with the display. Therefore, they considered this behavior as not safety-critical. 

\subsection{Benefits and Use Cases}
In general, participants agree, that the presented visualizations would benefit multiple use cases in the UX design process. Participants' statements describe that the three visualizations have great value for efficiently visualizing large amounts of interaction data and that they currently miss such possibilities in their daily work. P3 concludes that \textit{``[a]ll the information that brings you closer to the context of the user while you are sitting in the office behind your screen is extremely valuable''}. 

\textbf{Task Level View.} The Task Level View is considered very useful by all participants. They, in particular, appreciated the simple and intuitive representation of user flows. This is also shown insofar as they had no problems answering Study Questions SQ1-SQ3. P3 was especially interested in flows \fix{that can be} considered conspicuous because \textit{``[y]ou can find issues where nobody would even think of doing a qualitative study because you did not even think of this behavior. But if 5\% of all people behave that way there must be a reason for it and it should be further investigated''}. \fix{P3 further} added that \textit{``[...] there could be so many feature improvements based on the issues detected using this view''}. Similarly, P2 adds that \textit{``[they] currently have a collection of questions from different UX designers within the company that could, probably, be answered with this kind of visualization''}. \fix{The interviewee} further describes that a data-driven platform similar to the proposed one could have great benefit not only for UX experts but also for management and product development.

\textbf{Flow Level View.} In general, the participants agree that the Flow Level View is helpful in the design process. P1 states that \textit{``[b]eing able to see statistics like the median and the distribution of the sequences makes this visualization valuable when comparing different flows''}. P4 argues that it would also be interesting to see how these graphs change over time when people get more familiar with the system: \textit{``How do these graphs look like for novice users and how do they look like for experts users?''}. \fix{Furthermore, P1} adds that this would benefit the assessment of intuitiveness and learnability. P3 states that the distribution of sequences over the time on task is from particular interest because \textit{``[...] if a lot of users are at the far end of the distribution it would mean that a lot of them might have problems with this flow and I would be interested in why it takes them such a long time to complete the task''}. \fix{P1} further elaborates that it would be helpful to see specific sequences for identified outliers since the time on task alone indicates critical behavior.
	\begin{figure*}
		\centering
		\subfloat{\includegraphics[width=\linewidth]{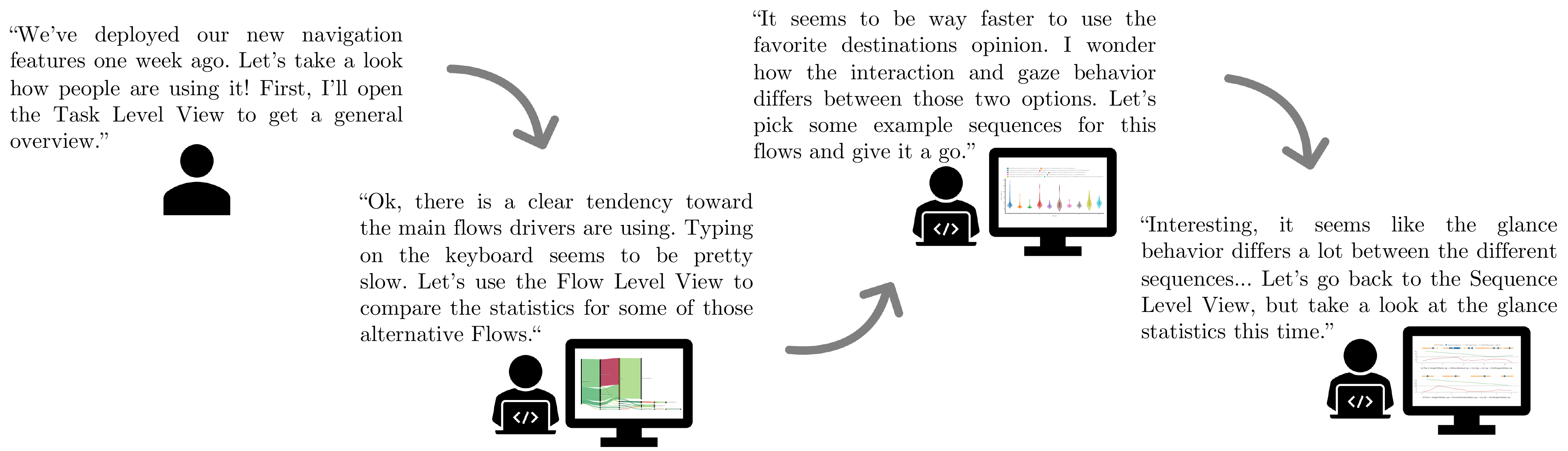}\label{fig:Feedback}}
		\caption{\fix{Exemplary Usage Scenario}}
		\label{fig:UsageGuide} 
	\end{figure*}

\textbf{Sequence Level View.}
All participants consider the Sequence Level View very helpful and argue that it plays an important role, especially in combination with the other visualizations. \fix{Whereas the other views present higher-level aggregated statistics, the visualization of specific sequences was helpful to develop a more precise understanding of how the interactions in the vehicle take place}. The additionally given information and especially the glance behavior data was considered very useful because \textit{ ``[o]ne can derive important information regarding the context to set the interaction into perspective''} (P4). Additionally, P3 emphasizes the importance regarding safety assessments because \textit{``it might be better to prioritize something slower but with fewer glances''}. P1 and P4 both explain that in order to get insights into glance behavior they, until now, had to set up specific lab studies.

\section{Conclusions}
This paper presents a Multi-Level User Behavior Analysis Framework providing insights into driver system interactions with \acp{IVIS} on three levels of granularity. The proposed approach consists of a telematics-based data logging and processing system, allowing live data collection from production vehicles. The presented visualizations are based on event sequence data, driving data, and glance behavior data. \fix{As a whole} they enable UX experts to quickly identify potential problems, quantify them, and examine their influence on glance or driving behavior using representative examples. \fix{An example that visualizes how the different views support each other and how UX experts may use them is given in Figure \ref{fig:UsageGuide}}.

The conducted user study shows that the presented visualizations help UX experts in designing \acp{IVIS}, assisting them in finding usability issues and unexpected user behavior. They report that they would use performance data more often if such visualizations would be available and argue that the generated insights would benefit the feature and requirements elicitation process. The Task Level View was considered the most helpful, closely followed by the Sequence Level View, followed by the Flow Level View. This coincides with the observations made during the evaluation study. 
\fix{During the study, participants switched between the different views depending on the type of information they were interested in. This consolidates our assumption that the different views support each other in a meaningful way and that different levels of detail are necessary to generate the best possible insights into driver \ac{IVIS} interaction.}

Our results show that visualizing large amounts of automotive interaction data using the proposed three visualizations is promising. However, we also identified points for improvement. One common suggestion is the mapping between user interactions and actual screens. This helps to interpret the visualizations without the need to know the names of the UI elements. Additionally, participants suggested making the visualizations visually more pleasing and proposed adding a dashboard-like overview of general statistics. \fix{This being a first exploratory approach, we only evaluated if participants interacted as intended and if they were able to generate the anticipated insights. For future iterations, it would be interesting to assess effectiveness and efficiency and compare multiple alternatives. Additionally, future evaluations should include participants outside of our research partner's organization.} None of our participants were affected by color vision deficiency, however, we have been advised to use a colorblind-friendly palette in future versions. 

Even if they do not directly influence the contribution of this work, ethical aspects of data collection, data security, and privacy are particularly important in the broader scope of this work. As of now, only company-internal testing vehicles contribute to the data collection. However, for future use cases, it is conceivable that customers contribute to the data collection and receive benefits such as earlier access to new features as compensation. The consent for data collection is given actively using the so-called ``opt-in'' standard. Therefore, users have full control over the decision whether or not to share their data to contribute to product improvement. As already mentioned, the data is completely anonymized, making it impossible to draw conclusions about individual users or their behavior.

By addressing various needs of automotive UX experts \cite{Ebel.2020a} the proposed approach is a first step toward better integration of quantitative user behavior data in automotive UX development. We envision the presented approach to be integrated into an overarching analysis platform allowing UX experts to freely explore large amounts of live data, collected from production or test vehicles to generate instant insights into in-car user behavior.

\bibliographystyle{ACM-Reference-Format}
\bibliography{VisualizationPaper}

\end{document}